\documentclass[prl,twocolumn, showpacs,preprintnumbers,amsmath,amssymb]{revtex4}

\usepackage{graphicx,color}
\usepackage{multirow}
\usepackage{braket}
\usepackage[colorlinks=true, urlcolor=blue, linkcolor = blue, citecolor = blue, pdfborder={0 0 0}]{hyperref}

\newcommand{\mns}{$\ket{-}\,$}
\newcommand{\double}{$\ket{\uparrow\downarrow}\,$}
\newcommand{\zero}{$\ket{0}\,$}
\newcommand{\sgma}{$\ket{\sigma}\,$}

\begin{document}

\title{Non-local spectroscopy of Andreev bound states}

\author{J.~Schindele}
\email{jens.schindele@unibas.ch}
\author{A.~Baumgartner}
\author{R.~Maurand}
\author{M.~Weiss}
\author{C.~Sch\"{o}nenberger}
\affiliation{Department of Physics, University of Basel, Klingelbergstrasse 82, CH-4056 Basel, Switzerland}

\date{\today}

\begin{abstract}
We experimentally investigate Andreev bound states (ABSs) in a carbon nanotube quantum dot (QD) connected to a superconducting Nb lead (S). A weakly coupled normal metal contact acts as a tunnel probe that measures the energy dispersion of the ABSs. Moreover we study the response of the ABS to non-local transport processes, namely Cooper pair splitting and elastic co-tunnelling, that are enabled by a second QD fabricated on the same nanotube on the opposite side of S. We find an appreciable non-local conductance with a rich structure, including a sign reversal at the ground state transition from the ABS singlet to a degenerate magnetic doublet. We describe our device by a simple rate equation model that captures the key features of our observations and demonstrates that the sign of the non-local conductance is a measure for the charge distribution of the ABS, given by the respective Bogoliubov-de Gennes amplitudes $u$ and $v$. 
\end{abstract}


\maketitle

\section{Introduction}
\begin{figure}[b]
	\centering
		\includegraphics{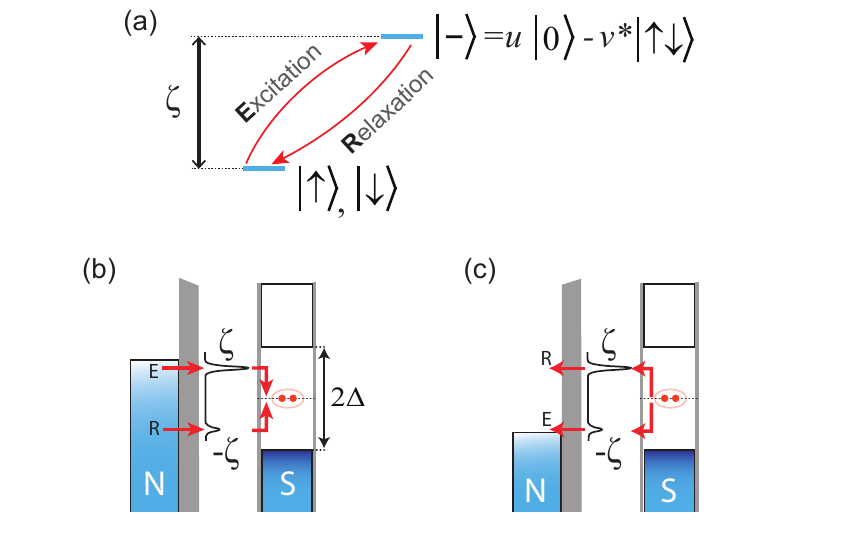}
	\caption{(a) Example of the low energy excitation spectrum of a QD-S device, with the magnetic doublet as GS, separated from the ABS by the Andreev addition energy $\zeta$. (b) Energy diagram of the local Andreev transport through a normal conducting tunnel probe. The alternation of an excitation and a relaxation process (labelled E and R) converts a normal current into a supercurrent. (c) Transport process at negative bias.}
	\label{fig:ES}
\end{figure}
In a Bardeen Cooper Schrieffer (BCS) superconductor the electrons gain a binding energy $2\Delta$ by pairing up in spin singlets known as Cooper pairs. This superconducting order can leak into non-superconducting materials placed in electrical contact with S. When this non-superconducting material is a quantum dot (QD) with a discrete energy spectrum, the proximity effect results in the formation of new sub-gap eigenstates named Andreev bound states (ABSs).
In a pictorial way one might think of the ABSs as emerging from the superposition of virtual Andreev reflections at the interface between the QD and a superconducting electrode (S). In each such Andreev reflection a Cooper pair (virtually) enters or leaves the QD, thereby mixing the even charge states of the QD. In the so-called superconducting atomic limit the ABS can be expressed as a BCS-like superposition of an empty and a doubly occupied QD level, denoted as $\ket{-}=u\ket{0}-v^{*}\ket{\uparrow \downarrow}$ \cite{Meng2009,Bauer2007,BraggioKoenig2011}. The \mns state is characterized by its energy $E_{-}$ and by the Bogoliubov-de Gennes (BdG) amplitudes $u$ and $v$. 
The odd charge states are not affected by the BCS condensate and remain eigenstates of the QD, forming a spin-degenerate doublet $\{\ket{\uparrow},\ket{\downarrow}\}$ \cite{Meng2009,Bauer2007,BraggioKoenig2011}. 

\subsection{Local spectroscopy of ABS }
The low energy excitation spectrum of a QD-S system is shown schematically in Fig.\ref{fig:ES}(a), where we chose the magnetic doublet to be the ground state (GS) and the ABS to be the excited state (ES). A natural experiment to measure the Andreev addition energy $\zeta=\vert E_{-}-E_{\uparrow,\downarrow}\vert$, defined as the energy difference between ABS and magnetic doublet, uses a normal conducting tunnel probe (N) in a N-QD-S geometry. If the tunnel coupling between N and the QD, $\Gamma_{\rm N}$, is sufficiently weak, the influence of the tunnel probe on the QD-S excitation spectrum is negligible and the differential conductance across the device, $G=\partial I/\partial V_{\rm SD}$, shows a peak for $\vert e V_{\rm SD}\vert=\zeta$ \cite{Tarucha2010,Pillet2010,DirksMason2011,Pillet2013,LeeDeFranceschi2013,KumarStrunk2013}. 

This peak in differential conductance represents the onset of a current through the Andreev channel when the electrochemical potential of the tunnel probe, $\mu_{\rm N}$, exceeds the addition energy, $\zeta$, as depicted in Fig.~\ref{fig:ES}(b). This allows an electron to tunnel across the barrier $\Gamma_{\rm N}$ and excite the QD, even in the presence of a large charging energy $U\gg\zeta$. The electron does not enter the \double state, but the \mns state, where the charge is shared between QD and S. The probability of this transition, $\ket{\uparrow}\xrightarrow{\smash{+1e}}\ket{-}$, scales with $v^{2}$, the weight of the \double term in the \mns state \citep{Wysokinski2012}. To relax back to the GS the QD takes up a second electron at negative energy $-\zeta$ from N, which is equivalent to the emission of a hole with energy $\zeta$ into N. The rate of this relaxation process is proportional to $u^{2}$, the probability to find the QD empty so that an electron can be added to reach the $\ket{\uparrow}$ state. A complete transport cycle, GS$\rightarrow$ES$\rightarrow$GS, reflects an incoming electron as a hole and transfers a Cooper pair to S with a probability proportional to $u^{2}v^{2}$. 

Since the \mns state is a superposition of an empty and a doubly occupied QD level, the same ES can be reached either by addition of an electron with positive energy $\zeta$ to the GS, or by removal of an electron with negative energy $-\zeta$ from the GS. Consequently the Andreev resonances are always observed symmetrically about Fermi level of the superconductor, which we define as reference potential $\mu_{\rm S}=0$. In case of a negative bias, $\mu_{\rm N}\leq-\zeta$, the QD is excited by removing an electron with negative energy $-\zeta$ from the QD and transferring it to N, as shown in Fig.~\ref{fig:ES}(c). The probability of this excitation, $\ket{\uparrow}\xrightarrow{\smash{-1e}}\ket{-}$, scales with $u^{2}$. Compared to the situation in Fig.~\ref{fig:ES}(b) the rates for excitation and relaxation are inverted and the direction of electron flow is reversed, but the Andreev current is again proportional to $u^{2}v^{2}$. Therefore local spectroscopy of ABS is not able to investigate the excitation and relaxation process individually in a controlled manner.

\subsection{Non-local spectroscopy of ABS} 
When a current is passed through the Andreev channel the QD fluctuates between $\{\ket{\uparrow},\ket{\downarrow}\}$ and \mns. In each such fluctuation the QD state changes between even and odd occupation, which requires the addition or removal of a single electron to the QD. If only local processes are considered the S-contact can not drive such transitions because the electrons at energies below $\Delta$ are paired and form a so-called BCS condensate. However, if a second QD is added to the QD-S system, higher order processes involving electrons from the second QD can deliver single electrons at sub-gap energies to one side of the superconductor.

\begin{figure}[t]
	\centering
		\includegraphics{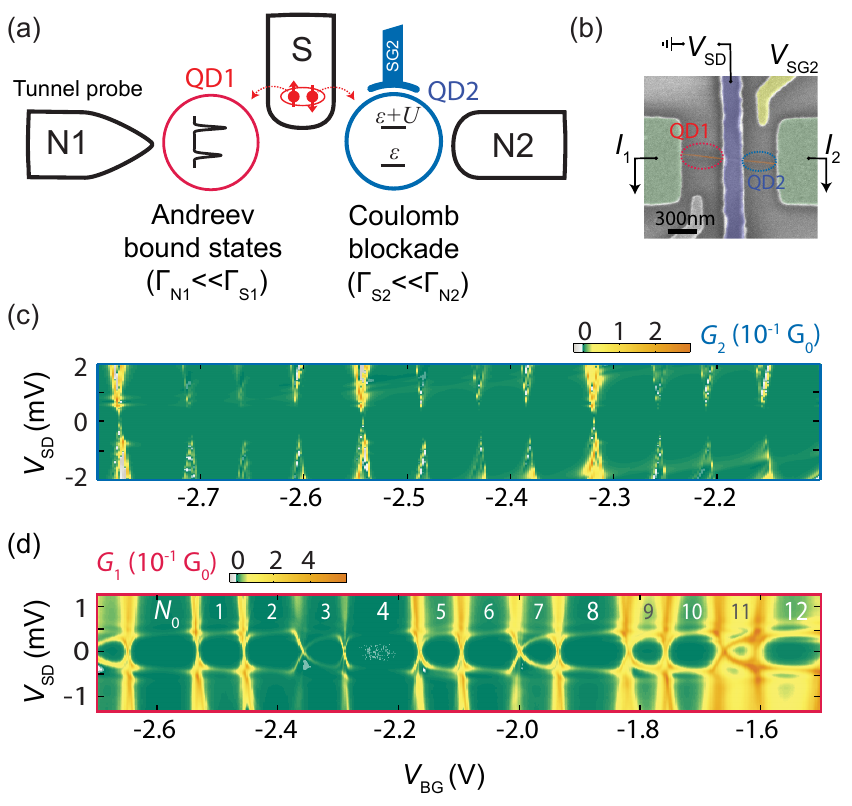}
	\caption{(a) Device schematic: two QDs are coupled to a common superconducting reservoir and two independent normal leads. The tunnel couplings follow $\Gamma_{\rm N1}\ll\Gamma_{\rm S1}$ and $\Gamma_{\rm S2}\ll\Gamma_{\rm N2}$. The subscript S/N labels the contact and the numbers refer to the respective QD. When both QDs are resonant Cooper pairs can split and leave S at different sites, thereby exiting the ABS on QD1. (b) Coloured SEM micrograph of the device and measurement set-up. (c,d) Differential conductances $G_{2}$ and $G_{1}$ as a function of the common source drain voltage, $V_{\rm SD}$, and back gate voltage, $V_{\rm BG}$.}
	\label{fig:Device}
\end{figure}

Figure \ref{fig:Device}(a) shows a sketch of the device geometry we consider. Two QDs (QD1 and QD2) are connected to two normal conducting drains (N1 and N2) and one common superconducting source. One possible process, in which the S-contact can excite QD1, is elastic co-tunnelling: an electron at energy $\zeta$ tunnels from QD2 to QD1 via a virtual quasiparticle state in S. Another mechanism is crossed Andreev reflection, also known as Copper pair splitting (CPS): a Cooper pair is coherently split into two electrons at opposite energies, here $\zeta$ and $-\zeta$, that leave S at different sites \cite{RecherLoss2001}. Recent experiments with similar device geometries demonstrated that the splitting of Cooper pairs can be controlled by tuning the levels of the individual QDs with local gates \cite{Hofstetter2009,Herrmann2010,Hofstetter2011,Schindele2012,DasHeiblum2012}. In contrast to these experiments we are able to resolve individual ABSs on one of the QDs. We then employ the CPS mechanism to excite these ABS. Thus, in our device the Cooper pairs play a twofold role. On the one hand, the Cooper pair condensate mixes the even charge states of QD1 as a result of the proximity effect. On the other hand, Cooper pairs can be split into individual charges that drive QD1 from even to odd occupation (or vice versa) with the assistance of QD2.

Since CPS and elastic co-tunnelling are coherent processes with electrons from two spatially separated QDs, we refer to them as non-local. In this paper we use local tunnelling spectroscopy to identify ABSs and then investigate the response of the ABS channel to non-local excitations. 
In section \ref{sec:Exp} we describe how the double QD device is realized with a carbon nanotube and present local and non-local transport measurements.
In section \ref{sec:Model} we introduce a simple rate equation model that explains our main experimental findings. We show that the non-local current reflects the relative amplitudes of the BdG amplitudes. 
In section \ref{sec:Conclusion} we summarize the results and conclude that non-local transport measurements provide a novel spectroscopic tool to investigate the charge distribution of the ABS -- an information that complements the knowledge of the Andreev addition energy $\zeta$ accessed by local tunnelling spectroscopy. 

\section{Experiment}
\label{sec:Exp}
\subsection{Device and measurement set-up} 
Figure \ref{fig:Device}(b) shows a colored scanning electron micrograph of our device and schematically the measurement set up. Two QDs are fabricated from a carbon nanotube (CNT) grown by chemical vapor deposition on a highly doped Si substrate capped with a 0.4$\,\rm \mu m$ insulating layer of thermal oxide. A Nb lead (50$\,\rm nm$ thick, 170$\,\rm nm$ wide), with a Ti contact layer (3$\,\rm nm$ thick) below, serves as superconducting reservoir. Together with two Ti/Au contacts ($5/50\, \rm nm$ thick) the S contact defines two QDs. The QDs can be tuned by applying a voltage $V_{\rm BG}$ to the Si substrate, which serves as global back gate, or by applying a voltage $V_{\rm SG2}$ to a local side gate in the vicinity of QD2. A second side gate near QD1 was not connected. We bias the device at S with $V_{\rm SD}$ and use two independent current voltage converters at N1 and N2 to obtain the currents through QD1 and QD2. The differential conductances through QD1, $G_{1}=\partial I_{1}/\partial V_{\rm SD}$, and through QD2, $G_{2}=\partial I_{2}/\partial V_{\rm SD}$, are measured simultaneously by standard lock-in technique, while varying the gate voltages and $V_{\rm SD}$. All measurements are carried out in a dilution refrigerator at a base temperature $T  \approx25\,\rm mK$.  

\subsection{Local transport measurements} 
The structure of the stability diagrams differs strongly for QD1 and QD2. The stability diagram of QD2 [Fig.~\ref{fig:Device}(c)] shows the well known pattern of Coulomb diamonds, disconnected by an induced transport gap of $2\Delta$ due to the superconductor, from which we extract $\Delta\approx 0.5\,\rm meV$. For voltages $\vert e V_{\rm SD} \vert < \Delta $ the conductance through QD2 is suppressed by a factor of $\sim10$. 

\begin{figure}[h]
	\centering
		\includegraphics{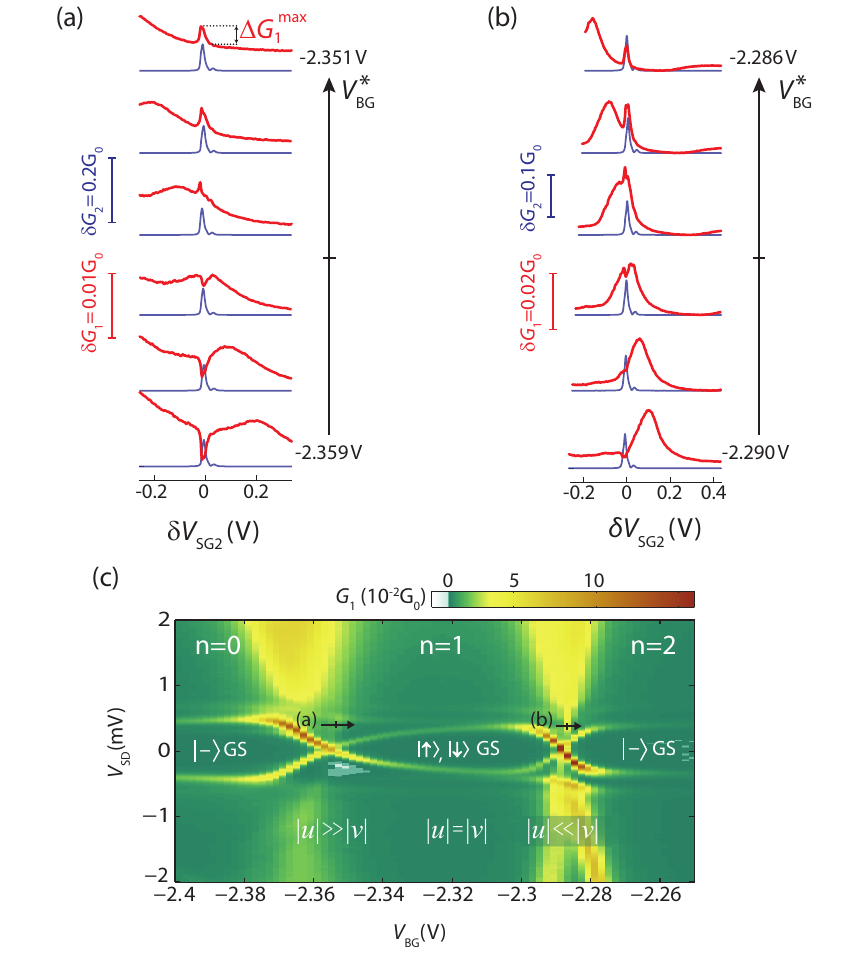}
 	\caption{(a,b) Simultaneously recorded differential conductances $G_{1}$ and $G_{2}$ as a function of $\delta V_{\rm SG2}=V_{\rm SG2}-\hat{V}_{\rm SG2}$ for increasing values of $V^{\text{*}}_{\rm BG}=V_{\rm BG}+\alpha \hat{V}_{\rm SG2}$. The source drain voltage was kept fixed at \mbox{$V_{\rm SD}=0.375\,\rm mV$}. The resonances in $G_{2}$ are accompanied by a non-local conductance change $\Delta G_{1}$ in $G_{1}$. (c) Stability diagram $G_{1}(V_{\rm SD},V_{\rm BG})$ for QD1 measured at $V_{\rm SG2}=0$. The black arrows indicate the direction along which the non-local signal is probed in (a,b). The sign change of $\Delta G_{1}$ coincides with the GS transitions of QD1.}
	\label{fig:slices}
\end{figure}

The conductance map for QD1 is shown in Fig.~\ref{fig:Device}(d) and in Fig.~\ref{fig:slices}(c), which zooms into the gate range around a diamond with odd occupation. Again the conductance is suppressed for $\vert V_{\rm SD} \vert <0.5\,\rm mV$, but inside the superconducting gap we observe two lines, positioned symmetrically about $V_{\rm SD}=0$, that cross each other near the diamonds edges. We interpret these sub-gap features as Andreev resonances at $\pm\zeta$. The crossing of two Andreev resonances at zero energy is associated with a quantum phase transition in which the GS of the QD changes from the \mns singlet to the magnetic doublet, or vice versa \citep{Bauer2007,Pillet2013,LeeDeFranceschi2013}. For odd occupation numbers the Coulomb repulsion, which favours the doublet GS, can prevail over the superconducting pairing, which favours the ABS as GS. At the phase boundary the energy of the \mns state equals the energy of the magnetic doublet and hence the Andreev resonances cross, i.e. $\pm\zeta=0$. For even occupation, where the QD is in the \mns GS, we find that the Andreev addition energy is pinned close to the gap edge, $\zeta\approx\Delta$.

Both QDs have similar charging energies of $\sim 5\, \rm meV$ and their stability diagrams exhibit a fourfold symmetry that is characteristic for clean CNT devices \cite{CobdenNydaard2002}. However, remaining disorder and spin orbit interactions lift the fourfold degeneracy, breaking up the CNT shells into two pairs of Kramer doublets \cite{Jespersen2011}. For QD1 we evaluate the separation between both Kramer doublets to be \mbox{$\delta=1\pm0.3\,\rm meV$}. Thus we treat the ABS as emerging from two-fold spin-degenerate energy levels, neglecting the influence of the additional orbital degree of freedom on the ABS spectrum.

\begin{figure}[h]
	\centering
		\includegraphics{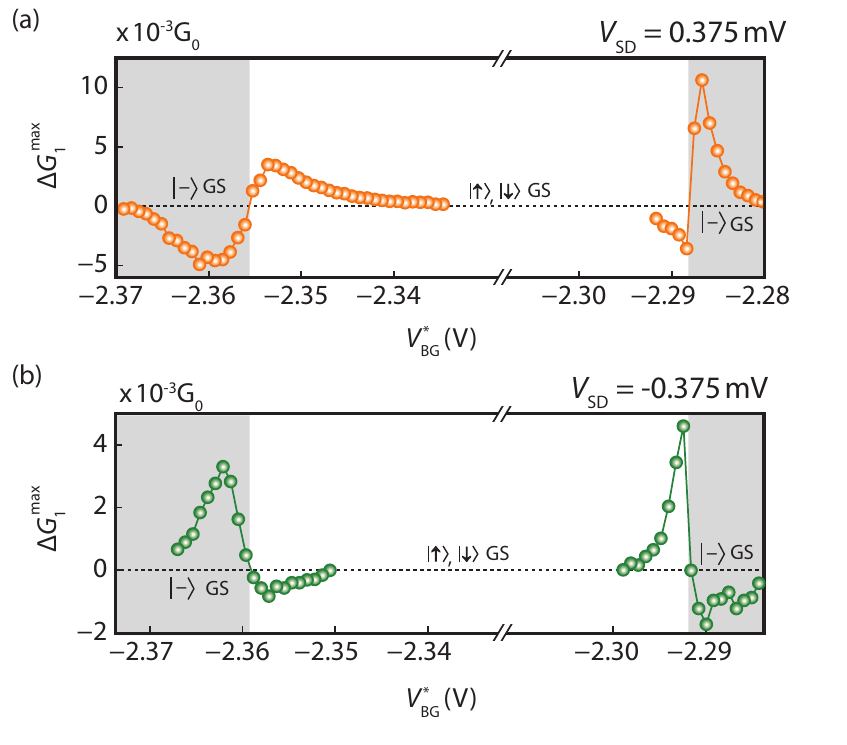}
 	\caption{Magnitude of the non-local differential conductance, $\Delta G_{1}^{\rm max}$, as a function of $V^{\text{*}}_{\rm BG}$, the back gate voltage corrected for the capacitive cross-talk from the side gate (see text). A bias of $V_{\rm SD}=0.375\,\rm mV$ (a) and  $V_{\rm SD}=-0.375\,\rm mV$ (b) was applied to the S contact.}
	\label{fig:DelG}
\end{figure}
\subsection{Non-local conductance correlations}
CPS and elastic co-tunnelling involve electron exchange with both QDs and can therefore be identified by studying correlations between the conductances $G_{1}$ and $G_{2}$. By tuning QD2 from Coulomb blockade to resonance, or vice versa, the non-local transport processes can be switched on and off, provided $\vert e V_{\rm SD}\vert \geq \vert \zeta \vert$. 
In Fig.~\ref{fig:slices}(a,b) we plot $G_{1}$ and $G_{2}$ as a function of the voltage applied to the local side gate at QD2. When a resonance of QD2 enters the bias window, which was set to $V_{\rm SD}=0.375\, \rm mV$, a sudden increase in the differential conductance $G_{2}$ is observed. These peaks in $G_{2}$ are accompanied by a conductance change $\Delta G_{1}$ in $G_{1}$. We ascribe these correlations $\Delta G_{1}(G_{2})$ to the non-local conductance caused by CPS and elastic co-tunnelling. To substantiate this interpretation we note that the conductance correlations tend to zero when superconductivity is suppressed, either by raising the temperature above $500\,\rm mK$ or by applying an external magnetic field $B_{\parallel} > 500\,\rm mT$ (see Appendix~\ref{sec:AppendixB}). 

By repeating these correlation measurements for many consecutive values of $V_{\rm BG}$ we can map out how the non-local signal depends on the energy level configuration of QD1. To correct for the capacitive cross-talk from the side gate to QD1 we introduce the new variable $V^{\text{*}}_{\rm BG}=V_{\rm BG}+\alpha \hat{V}_{\rm SG2}$. Here $\hat{V}_{\rm SG2}$ is the side gate voltage for which the non-local conductance takes its maximal value, $\Delta G_{1}^{\rm max}$, and $\alpha = 1.56\times 10^{-2}$ is a geometry dependent factor that accounts for the respective gate efficiency. The variable $V^{\text{*}}_{\rm BG}$ allows to assign a position in the stability diagrams of QD1, measured at $V_{\rm SG2}=0$, to the non local signals, measured at $\hat{V}_{\rm SG2}\neq 0$. 
In Fig.~\ref{fig:slices}(c) we indicate the direction along which $\Delta G_{1}^{\rm max}$ is probed in Fig.~\ref{fig:slices}(a,b) by black arrows. The conductance correlations can be either positive or negative, i.e. $G_{1}$ can show a peak or a dip at the QD2 resonance, depending on $V^{\text{*}}_{\rm BG}$. Strikingly, the turnover from a negative to a positive non-local conductance coincides with the quantum phase transition in which the GS changes from the ABS singlet to the magnetic doublet. 

In Fig.~\ref{fig:DelG} we plot the evolution of $\Delta G_{1}^{\rm max}$ over the complete back gate range of a odd QD1 state for opposite bias voltages $V_{\rm SD}=\pm 0.375\,\rm mV$. Starting from the left side of Fig.~\ref{fig:DelG}(a) a negative non-local signal starts to build up when the Andreev resonance enters the bias window, $\zeta<\vert e \vert V_{\rm SD}=0.375\,\rm mV$, at $V^{\text{*}}_{\rm BG}\approx -2.37\,\rm V$. The magnitude of $\Delta G_{1}^{\rm max}$ increases towards the singlet--doublet phase boundary where it rapidly changes sign. In the doublet GS region the positive correlations decay and become immeasurably small around the centre of the plot. As the right GS transition is approached the non-local signal builds up again, but with a negative sign. Around $V^{\text{*}}_{\rm BG}\approx -2.288\,\rm V$, where we expect the \mns state to become the GS, the sign of $\Delta G_{1}^{\rm max}$ is again inverted. The evolution of the non-local signal at a negative bias voltage of $V_{\rm SD}=-0.375\,\rm mV$, shown in Fig~\ref{fig:DelG}(b), exhibits a similar behaviour, except for a sign change that results from the reversal of the bias voltage. 

Comparing the left and the right side of Fig.~\ref{fig:DelG}(a,b) we notice a sharper reversal of $\Delta G_{1}^{\rm max}$ at the right GS transition.  
However, the slope of the dispersion $\zeta(V_{\rm BG})$ in Fig.~\ref{fig:slices}(c) is also steeper at the right GS transition, implying a more rapid crossover between different GSs than for the left GS transition. We speculate that this asymmetry results from a gate dependence of $\Gamma_{\rm S1}$, which decreases for increasing  $V_{\rm BG}$.

The sign change of the non-local signal is reminiscent of the 0--$\pi$ transition in S-QD-S Josephson junctions. There, a reversal of the supercurrent across the device is observed when the GS of the QD changes from singlet to doublet \cite{vanDamKouwenhoven2006,Jorgensen2007,Pillet2010, Maurand2012}. However, the back gate evolution of $\Delta G_{1}^{\rm max}$ demonstrates that the sign of the non-local signal is not merely determined by the GS of QD1, but also changes in the doublet GS region and under reversal of bias voltage. Hence, the sign of $\Delta G_{1}$ can not be explained by analogy to the supercurrent reversal at the 0--$\pi$ transition.

\section{Model}
\label{sec:Model}
\begin{figure}[t]
	\centering
		\includegraphics{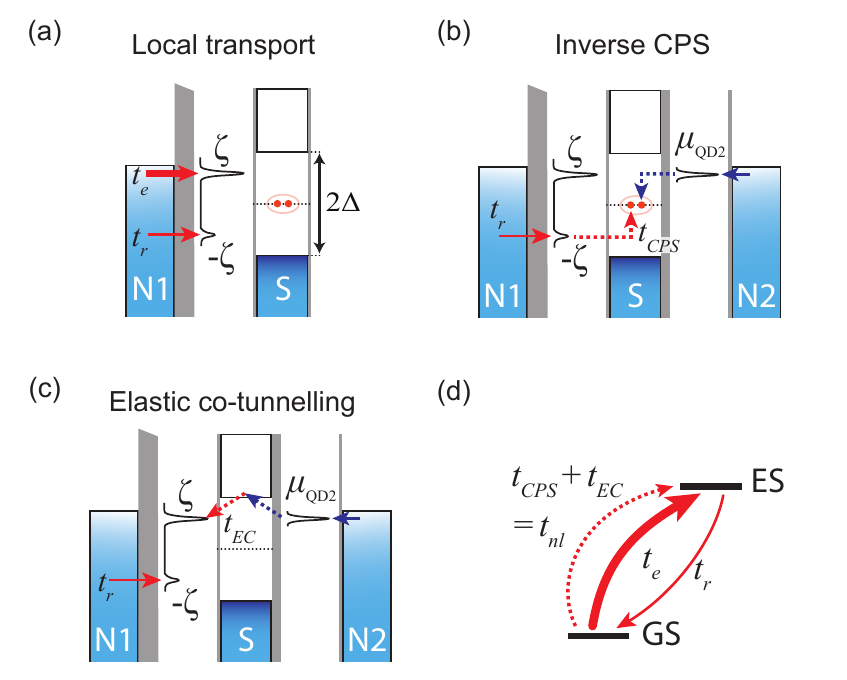}
 	\caption{(a-c) Energy diagrams of the local and non-local transport processes considered in our model. (d) Two-level rate equation model model for QD1. The non-local processes with rate $t_{n\ell}$ (red dashed arrow) only change the occupation probabilities but do not contribute to the current through N1.}
	\label{fig:Model}
\end{figure}

To understand the nature of the observed non-local signals we discuss the relevant transport processes and their impact on the conductance $G_{1}$. Assuming $\vert e \vert V_{\rm SD}>\zeta$, the local Andreev channel gives rise to a background current that flows from N1 to S, as shown in Fig.~\ref{fig:Model}(a), where $t_{e}$ and $t_{r}$ denote the rate of the local excitation and the local relaxation by electrons from N1. If QD2 is tuned into resonance it can provide single electrons with energy $\zeta$. This configuration allows the non-local creation of Cooper pairs in a process inverse to CPS: an electron from QD2 with energy $\zeta$ and an electron from QD1 with energy $-\zeta$ pair up and enter S in a distance on the order of the superconducting coherence length [Fig.~\ref{fig:Model}(b)]. We refer to the rate of this process as $t_{CPS}$. In addition, an electron from QD2 can also co-tunnel via a quasiparticle state in S and excite QD1, as shown in Fig.~\ref{fig:Model}(c), where we define the rate of elastic co-tunnelling as $t_{EC}$. 

We note that non-local relaxation processes, which require that QD2 absorbs electrons at energy $\zeta$, are suppressed by the coupling asymmetry of QD2: The condition $\Gamma_{\rm S2}\ll\Gamma_{\rm N2}$ implies that QD2 is refilled much faster from N2 than from S. Therefore, the relaxation of QD1 is dominated by the same local process, independent of the nature of the preceding excitation.

\subsection{Rate equation}
To model the conductance through QD1 we formulate a two-level rate equation, graphically illustrated in Fig.~\ref{fig:Model}(d). The steady state occupation probabilities of GS and ES, denoted $P_{\rm GS}$ and $P_{\rm ES}=1-P_{\rm GS}$, are given by
\begin{equation}
\frac{d}{dt}P_{\rm ES}=\left(
t_{e}+t_{n\ell}\right) P_{\rm GS}
-t_{r} P_{\rm ES}=0,
\label{eq:SteadyState}
\end{equation}
where $t_{n\ell}=t_{CPS}+t_{EC}$ is the sum of both non-local excitation rates. 

From the occupation probabilities one can calculate the current through the tunnel probe N1 
\begin{equation}
I_{1}=\frac{e}{\hbar}\left(t_{e}P_{\rm GS}+t_{r}P_{\rm ES}\right).
\label{eq:G1}
\end{equation}
The influence of $t_{n\ell}$ on $I_{1}$ is hidden in the occupation probabilities $P_{\rm GS}$ and $P_{\rm ES}$, which are modified according to Eq.~(\ref{eq:SteadyState}) when $t_{n\ell}$ changes. The non-local excitations, depicted by the dashed arrows in Fig.~\ref{fig:Model}(b,c), do not exchange electrons with N1. Hence, the current through the barrier $\Gamma_{\rm N1}$ is only carried by local excitation and relaxation processes. In the absence of non-local transport Eq.~(\ref{eq:G1}) simplifies to $I_{1}(t_{n\ell}=0)=\frac{e}{\hbar}\frac{2t_{e}t_{r}}{t_{e}+t_{r}}$. To calculate the non-local current we subtract this local background from the total current, which yields 
\begin{eqnarray}
	\begin{split}
\Delta I_{1} 	 &= I_{1}(t_{n\ell}\neq 0)-I_{1}(t_{n\ell}= 0)\\
				 &= \frac{e}{\hbar} t_{n\ell} P_{\rm GS} \frac{t_{r}-t_{e}}{t_{r}+t_{e}}.	
	\end{split}
	\label{eq:DeltaG1}
\end{eqnarray}
As one may expect, the non-local current is proportional to the excitation rate $t_{n\ell}$ and the occupation probability of the GS, $P_{\rm GS}=t_{r}/(t_{e}+t_{r}+t_{n\ell})$. However, the sign of $\Delta I_{1}$ is determined by $t_{r}-t_{e}$, the relative strength of the local rates $t_{e}$ and $t_{r}$. This can be understood by considering a very asymmetric situation, $t_{r}\ll t_{e}$, as assumed in Fig.~\ref{fig:Model}. $t_{r}$ limits the current and the QD is "stuck" in the ES most of the time. The non-local processes increase this imbalance, but without contributing to the current through N1. The QD gets even more "stuck" in the ES and the current flow is hindered, $\Delta I_{1}<0$. In the reversed situation, $t_{r}\gg t_{e}$, the excitation rate is the bottleneck. Here the non-local excitations bypass this bottleneck, leading to an increased current, $\Delta I_{1}>0$. When the asymmetry between $t_{e}$ and $t_{r}$ decreases the sign of $\Delta I_{1}$ remains the same, but the non-local current also decreases and finally vanishes for $t_{e}=t_{r}$. 
 
The gate evolution of the rates $t_{e}$ and $t_{r}$ is determined by the physics of ABSs. We first discuss these rates in the limit $\Delta\rightarrow\infty$, where analytic expressions for the eigenstates of the QD-S system can be found. Later we compare these results to numerical calculations from the literature that consider a finite gap and therefore represent a more realistic scenario.   

Figure~\ref{fig:ModelResults}(a) shows the dispersion relation of the Andreev resonance in the limit $\Delta\rightarrow\infty$ calculated with the analytic expressions given in \citep{Meng2009} for $\Gamma_{\rm S1}=0.37$ in dimensionless energy units. The energy level of the QD, $\epsilon_{\rm d}$, is parametrized by $\delta=\epsilon_{\rm d}+U/2$. The local transport rates can be calculated with Fermi's golden rule \citep{BraggioKoenig2011,Wysokinski2012}, which yields  
\begin{equation}
	\begin{split}
&\ket{\sigma}\xrightarrow{+1e}\ket{-}:\; t_{e}^{+}=
\Gamma_{\rm N1}   
\underbrace{\vert\braket{-|d_{\bar{\sigma}}^{\dagger}|\sigma}\vert^{2}}         _{v^{2}} 
f_{1}(\zeta)
\\
&\ket{-}\xrightarrow{+1e}\ket{\sigma}:\; t_{r}^{+}=
\Gamma_{\rm N1}
\underbrace{\vert\braket{\sigma|d_{\sigma}^{\dagger}|-}\vert^{2}}           _{u^{2}}
f_{1}(-\zeta)
	\end{split}
\label{eq:t+}
\end{equation}
and
\begin{equation}
	\begin{split}
&\ket{\sigma}\xrightarrow{-1e}\ket{-}:\; t_{e}^{-}=
\Gamma_{\rm N1}   
\underbrace{\vert\braket{-|d_{\sigma}|\sigma}\vert^{2}}         _{u^{2}} 
(1-f_{1}(\zeta))
\\
&\ket{-}\xrightarrow{-1e}\ket{\sigma}:\; t_{r}^{-}=
\Gamma_{\rm N1}
\underbrace{\vert\braket{\sigma|d_{\bar{\sigma}}|-}\vert^{2}}           _{v^{2}}
(1-f_{1}(-\zeta)).
	\end{split}
\label{eq:t-}
\end{equation}
Here $f_{1}(E)$ is the Fermi function of the lead N1, $d_{\sigma}^{\dagger}$ ($d_{\sigma}$) is the creation (annihilation) operator of QD1 for an electron with spin $\sigma=\uparrow,\downarrow$ and $\bar{\sigma}$ denotes the spin opposite to $\sigma$. For a sufficiently positive bias $\vert e\vert V_{\rm SD}>\zeta$ we can approximate $f_{1}(\pm\zeta)\approx 1$, hence the rates $t^{-}_{e}$ and $t^{-}_{r}$ can be neglected. In Fig.~\ref{fig:ModelResults}(c) we plot the rates $t^{+}_{e}$ and $t^{+}_{r}$which reflect the evolution of the BdG amplitudes $v^{2}$ and $u^{2}$ with the QD energy. When the GS changes the initial and final state of the respective matrix elements are interchanged and the rates $t^{+}_{e}$ and $t^{+}_{r}$ are inverted. 

The non-local excitation rate relevant for positive bias can be written as
\begin{equation}
\begin{split}
\ket{\sigma}\rightarrow\ket{-}:\; t_{n\ell}^{+}=
	&\left( k_{\rm CPS}  \vert\braket{-|d_{\sigma}|\sigma}\vert^{2} 
	+k_{\rm EC}  \vert\braket{-|d_{\bar{\sigma}}^{\dagger}|\sigma}\vert^{2}\right)\\
&\times\varrho_{\rm QD2}(\zeta) f_{2}(\zeta),
\end{split}
\label{eq:t_nl}
\end{equation}
where $\varrho_{\rm QD2}(E)$ is the spectral density of QD2, $f_{2}(E)$ is the Fermi function of the lead N2 and the pre-factors $k_{\rm CPS}$ and $k_{\rm EC}$ give the respective process efficiencies. The rate $t_{n\ell}^{-}$ can be obtained by the following replacements: \mbox{$d_{\sigma} \leftrightarrow d_{\bar{\sigma}}^{\dagger}$}, $\zeta \rightarrow -\zeta$ and $f_{2}\rightarrow 1-f_{2}$. 

\subsection{Model results and comparison with experiment}
\begin{figure}[t]
	\centering
		\includegraphics{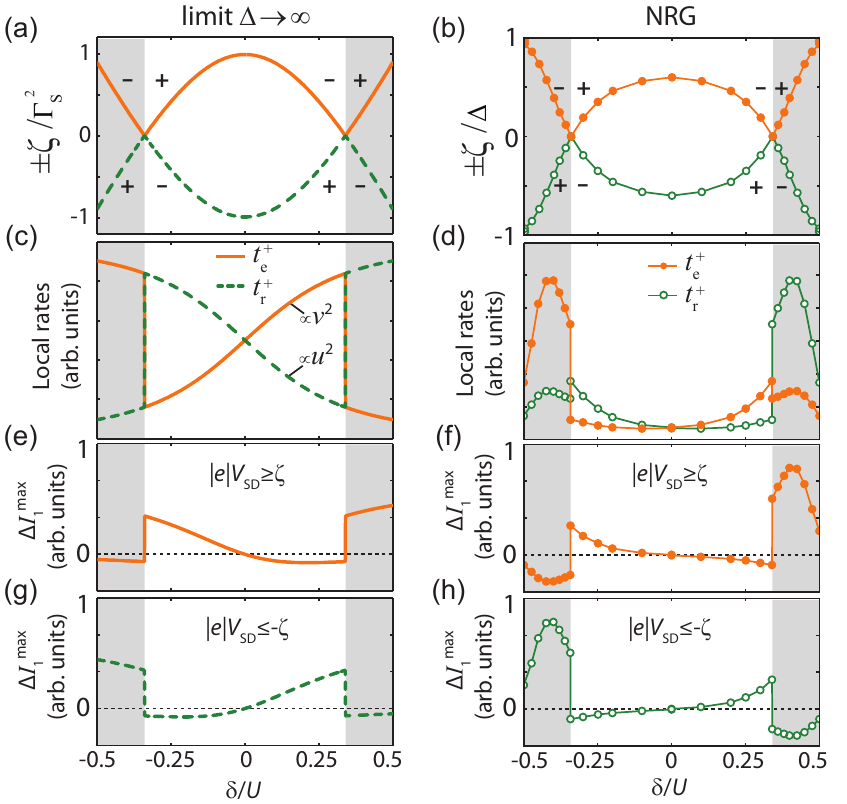}
 	\caption{Model for non-local signals calculated in the limit $\Delta\rightarrow\infty$ for $\Gamma_{\rm S1}=0.37$ (left column) and based on NRG results for the parameters $U=1$, $\Delta=0.01$ and $U/\pi\Gamma_{S}=5$ extracted from reference \citep{Bauer2007} (right column). The QD energy is parametrized by $\delta=\epsilon_{\rm d}+U/2$. The grey shaded regions indicate the \mns GS. (a,b) Dispersion of the Andreev resonances. The plus and minus symbols denote the sign of the non-local current in the respective region. (c,d) Local rates for $\vert e\vert V_{\rm SD}\geq\zeta$. (e,f) Non-local current $\Delta I_{1}$ for $\vert e\vert V_{\rm SD}\geq\zeta$ and (g,h) for \mbox{$\vert e\vert V_{\rm SD}\leq-\zeta$}.}
	\label{fig:ModelResults}
\end{figure} 

In Fig.~\ref{fig:ModelResults}(e,g) we plot the gate evolution of the maximum non-local current for positive and negative bias, calculated from Eq.~(\ref{eq:DeltaG1}) with the assumption $k_{\rm EC}=k_{\rm CPS}$. Despite the oversimplification $\Delta\rightarrow\infty$ the model captures the main features of our experimental findings. The sign of the non-local current alternates in the same order as in the experiment (see Fig.~\ref{fig:DelG}), going through two sharp transitions and one smooth transition. 

The sharp reversal of $\Delta I_{1}$ is the signature of the GS transition, in which the rates $t_{e}$ and $t_{r}$ are inverted. In the experiment this transition is smeared out by the broadening of the Andreev resonance not considered in our model. In the doublet GS the non-local conductance  changes gradually from positive to negative values, owing to the gate evolution of the BdG amplitudes. As the dot energy is increased the weight of the \mns state is shifted from the \zero -term to the \double -term, thereby gradually moving the average location of the two electron charges confined in the ABS from the superconductor to the QD. This continuous change of the BdG amplitudes leads to a smooth reversal of the non-local current at the electron-hole (e-h) symmetry point ($\delta=0$), where $t_{e}=t_{r}$.    

In case of a finite superconducting gap exchange interactions between the \sgma state and quasiparticles can lead to a spin screening of the \sgma state. This Kondo effect complicates the theoretical treatment of the problem. The wavefunction of the doublet state aquires a singlet admixture and the dispersion relation of the Andreev resonances, as well as the transport rates become renormalised. An analytical solution of the problem is not possible, but the numerical renormalization group method (NRG) provides a reliable approach to calculate the QD spectral densities \cite{LevyYeyati2011}.

In the right column of Fig.~\ref{fig:ModelResults} we test our model with NRG results calculated in reference \cite{Bauer2007} for the parameters $U=1$, $\Delta=0.01$ and $U/\pi\Gamma_{S}=5$. The dispersion of the Andreev resonances for these parameters, shown in Fig.~\ref{fig:ModelResults}(b), resemble our experiment. The local transport rates, plotted in Fig.~\ref{fig:ModelResults}(d), are given by the spectral weight of the respective Andreev resonance. To calculate the non-local current we assume again $k_{\rm CPS}=k_{\rm EC}$. Figure~\ref{fig:ModelResults}(f,h) shows that the qualitative behaviour of the non-local signal is altered only marginally when interactions with quasiparticles are considered. The main effect of the finite gap on our model originates from a suppression of the local transport rates when the Andreev resonance approaches the gap edge, i.e. $t_{e/r}\rightarrow0$ for $\zeta\rightarrow\Delta$. This leads to a cut-off of the non-local signal at the ends of the inspected gate range ($\delta=\pm 0.5$) and a more rapid decay towards the e-h symmetry point compared to the $\Delta\rightarrow\infty$ case. Both of these modifications in the line shape of the non-local signal improve the agreement with our experimental findings in Fig.~\ref{fig:DelG}.

\section{Discussion and Conclusion}
\label{sec:Conclusion}
We experimentally investigated a CNT QD, strongly coupled to a superconducting niobium lead. By local transport spectroscopy through a normal conducting tunnel probe we could resolve individual ABSs in the excitation spectrum of the QD-S system. A second QD, coupled parallel to the same S-contact, allowed to excite these ABSs also by non-local processes, namely CPS and elastic co-tunnelling. We found appreciable non-local correlations in the conductance through both QDs. These non-local signals change sign with reversed bias and exhibit a complex gate dependence with a sign change at the GS transition and a sign change when the e-h symmetric point is crossed. We qualitatively explain this rich behaviour in a simple rate equation model.

In our model the sign of the non-local current is determined by the asymmetry between the local excitation and relaxation rates. In the limit $\Delta\rightarrow\infty$ this asymmetry is given by the difference of the BdG amplitudes, $\gamma(v^{2}- u^{2})$, where the pre-factor $\gamma=\pm1$ changes sign when the GS or the bias direction changes. One can ascribe a physical meaning to this term by rewriting it as $2v^{2}-1$, using the normalization condition $u^{2}+v^{2}=1$. Multiplying with the electron charge, this corresponds to the charge difference between ES and GS,
\begin{equation}
\Delta Q=Q_{\rm ES}-Q_{\rm GS},
\end{equation}
where the average on-site charge in the \mns state is given by the expectation value of the number operator, $Q_{-}=e\braket{-|\sum_{\sigma}d^{\dagger}_{\sigma}d_{\sigma}|-}=2e\,v^{2}$. The QD charge in the doublet state is $Q_{\sigma}=1e$.

While local spectroscopy measures the energy difference between the ES and the GS, $\zeta=E_{\rm ES}-E_{\rm GS}$, the non-local signals provide a spectroscopic tool to investigate the charge difference between both states. However, a quantitative determination of $\Delta Q$ is hindered by the lack of knowledge about $t_{n\ell}$. Still we are able to \textit{qualitatively} follow the gate evolution of $\Delta Q$, which is a direct witness of the competition between repulsive Coulomb interactions and the superconducting pairing, associated with an attractive electron-electron interaction.  

The \mns state, being subject to quantum fluctuations of the charge, allows continuous changes of the mean QD charge. We were able to indirectly observe this gradual charging of the ABS by following the smooth crossover from a positive to a negative non-local signal when the QD is in the doublet GS. When $\Delta Q$ becomes negative, the QD holds more charge in the GS than in the first ES -- a situation that can only occur in the presence of attractive interactions. At the GS transition, which is identified by the continuous crossing of the two Andreev resonances in local spectroscopy, the sign of $\Delta Q$ is inverted. The resulting abrupt reversal of the non-local current constitutes a novel experimental probe of the discontinuity characteristic for such quantum phase transitions. 

In conclusion, we established a new spectroscopy method to study ABSs in QDs. Our method complements local tunnelling spectroscopy and indirectly provides access to the BdG amplitudes of the ABSs, yielding a novel experimental view on the superconducting proximity effect in QDs.

\section{Acknowledgements}
J.S. would like to thank Doroth\'{e}e Hug and Stefan Nau for advice during the sample fabrication. We acknowledge financial support by the EU FP7
Project SE$^{2}$ND, the EU ERC Project QUEST, the Swiss Science Foundation SNF including the NCCR-QSIT and the NCCR-Nano.

\appendix
\section{Bias dependence}
\label{sec:AppendixA}
\begin{figure}[b]
	\centering
		\includegraphics{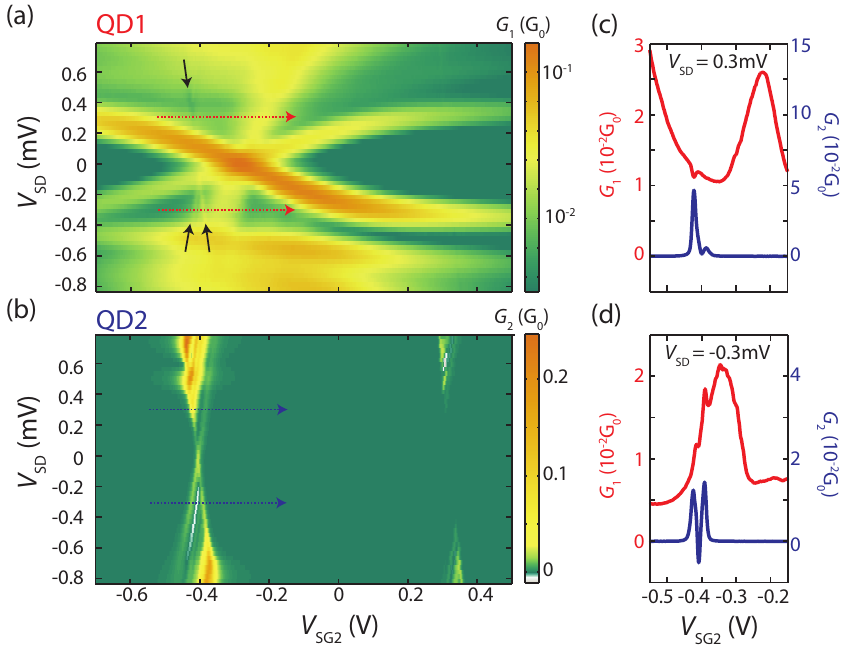}
 	\caption{(a,b) Simultaneously recorded differential conductances $G_{1}$ and $G_{2}$ as a function of side gate voltage, $V_{\rm SG2}$, and source drain bias, $V_{\rm SD}$, at $V_{\rm BG}=-2.284\,\rm V$. The black arrows in (a) guide the eye to conductance correlations $\Delta G_{1}(G_{2})$. (c) Cross sections of $G_{1}$ and $G_{2}$ for $V_{\rm SD}=+0.3\,\rm mV$. The large peak in $G_{2}$ correlates with a dip in $G_{2}$, i.e. the non-local signal $\Delta G_{1}(G_{2})$ is negative. (d) Cross section for $V_{\rm SD}=-0.3\,\rm mV$, yielding positive conductance correlations.}
	\label{fig:FBCPS}
\end{figure}

Figure~\ref{fig:FBCPS}(a,b) shows the simultaneously recorded differential conductances $G_{1}$ and $G_{2}$ as a function of $V_{\rm SG2}$ and $V_{\rm SD}$ at $V_{\rm BG}=-2.284\,\rm V$. The lever-arm of $V_{\rm SG2}$ to QD1 is about 8 times weaker than to QD2. Therefore the Andreev resonances in Fig.~\ref{fig:FBCPS}(a) appear very broad and smeared out compared to the Coulomb diamonds in Fig.~\ref{fig:FBCPS}(b). This separation of energy scales makes it easy to identify conductance correlations $\Delta G_{1}(G_{2})$, e.g. the ones indicated by the black arrows in Fig.~\ref{fig:FBCPS}(a), where a shallow imprint of the left diamond from Fig.~\ref{fig:FBCPS}(b) is observed. Fig.~\ref{fig:FBCPS}(c,d) show cross sections at constant bias voltages that demonstrate the sign reversal of $\Delta G_{1}(G_{2})$ with opposite bias. We note that otherwise the bias dependence of the non-local conductance is surprisingly weak. The intensity of the non-local conductance line is approximately constant between the Andreev resonance and the gap edge. Another intriguing feature in Fig.\ref{fig:FBCPS} is the slightly tilted vertical line, running exactly through the crossing point of the Andreev resonances, $\pm \zeta=0$. Such lines, also visible in the data from reference \citep{Pillet2010}, may be explained as follows. In the region $\vert \zeta \vert \leq \vert e V_{\rm SD} \vert \leq \vert \Delta \vert$ the Andreev resonance is the only conductance channel and the local current through the device is constant. The two Andreev resonances, $\zeta$ and $-\zeta$, have different conductances. When the two resonances cross, the current through the Andreev channel changes as a step function, yielding a peak in differential conductance. Thus, this line can be interpreted as a finite bias signature of the GS transition. Its slope is given by the capacitive cross-talk from the source contact.
However, the reason for the conductance difference between $\zeta$ and $-\zeta$, also observed in \citep{Tarucha2010,DirksMason2011,LeeDeFranceschi2013,Pillet2013,Pillet2010}, remains unclear. One possible explanation might be a soft superconducting gap for which quasiparticle states at energies $E<\Delta$ are available \cite{TakeiDasSarma2013}. This scenario would also allow tunnelling processes that break the e-h symmetry of the local sub-gap transport, e.g. the tunnelling of an electron from N1 to QD1 to a quasiparticle state in S. In this case the complete transport cycle, GS$\rightarrow$ES$\rightarrow$GS, has a probability proportional to either $v^{4}$ or $u^{4}$. 

\begin{figure}[t]
	\centering
		\includegraphics{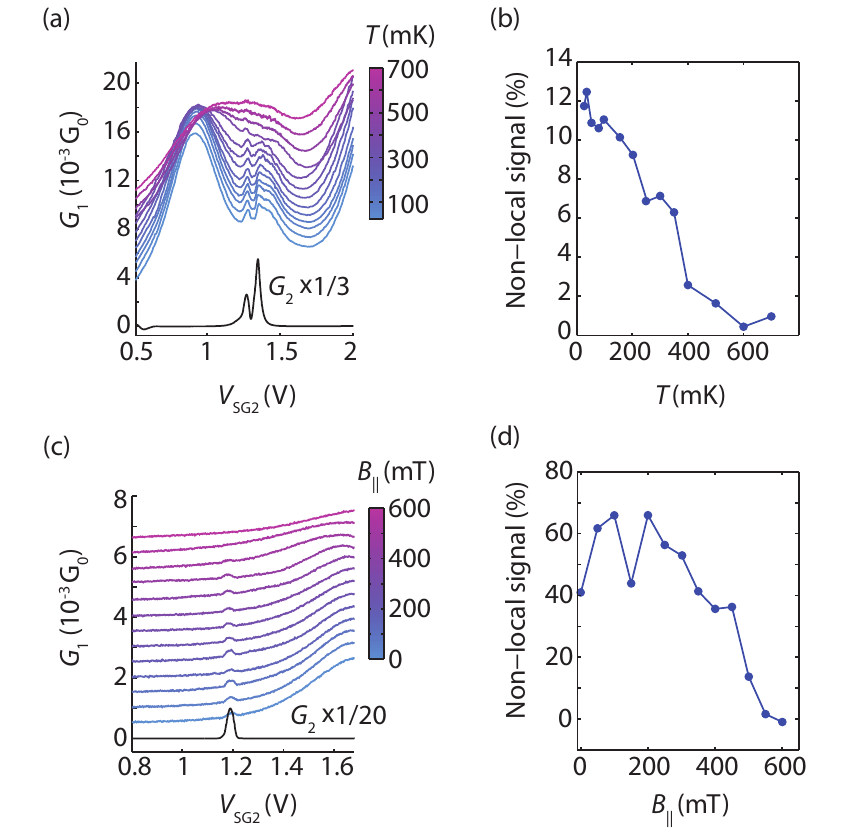}
 	\caption{(a) Simultaneously recorded $G_{1}$ and $G_{2}$ as a function of $V_{\rm SG2}$ at $V_{\rm BG}=-1.958\,\rm V$ and $V_{\rm SD}=-0.45\,\rm mV$. The non-local conductance variations $\Delta G_{1}$ tend to zero as the base temperature is increased. (b) Visibility of the non-local signal $\Delta G_{1}/G_{1}$ vs. $T$ obtained from the data shown in (a). (c) Simultaneous recorded $G_{1}$ and $G_{2}$ as a function of $V_{\rm SG2}$ at $V_{\rm BG}=15.583\,\rm V$ and $V_{\rm SD}=0\,\rm mV$. The non-local conductance variations $\Delta G_{1}$ tend to zero when an external magnetic field is applied. The field direction is parallel to the plane of the S-contact. (d) Visibility of the non-local signal $\Delta G_{1}/G_{1}$ vs. $B_{\vert\vert}$ obtained from the data shown in (c).}
	\label{fig:TandB}
\end{figure}
\section{Temperature and magnetic field dependence}
\label{sec:AppendixB}
Figures~\ref{fig:TandB}(a) and \ref{fig:TandB}(c) show the temperature and magnetic field dependence of the non-local conductance. Figures~\ref{fig:TandB}(b) and \ref{fig:TandB}(d) plot the visibility of the non local signal, i.e. $\Delta G_{1}^{\rm max}/G_{1}$, where $G_{1}$ is the local background conductance \cite{Schindele2012}. The non-local conductance decreases when superconductivity is suppressed and vanishes around a temperature of $\sim 500\,\rm mK$ or an in plane magnetic field of $\sim 500\,\rm mT$. These measurements demonstrate that the conductance correlations are mediated by superconductivity.


\end{document}